\documentclass{article}
\usepackage{PRIMEarxiv}

\usepackage[utf8]{inputenc} % allow utf-8 input
\usepackage[T1]{fontenc}    % use 8-bit T1 fonts
\usepackage{hyperref}       % hyperlinks
\usepackage{url}            % simple URL typesetting
\usepackage{booktabs}       % professional-quality tables
\usepackage{amsfonts}       % blackboard math symbols
\usepackage{nicefrac}       % compact symbols for 1/2, etc.
\usepackage{microtype}      % microtypography
\usepackage{fancyhdr}       % header
\usepackage{graphicx}       % graphics
\usepackage{xcolor,colortbl}
\usepackage{lscape}

\title{Statistical analysis on the effectiveness of a low insulin index, alkaline and functional diet evaluated with machine learning techniques
}

\author{
F. A. Conventi\\
Università degli studi di Napoli "Parthenope" \\
Naples, Italy\\
\And 
F. Cirotto, A. D'Avanzo, Ag. De Iorio, E. Rossi\\
Università degli studi di Napoli "Federico II"\\
Dipartimento di Fisica\\
Naples, Italy 
\And
An. De Iorio, M. Forte, F. Miele, G. Perna, B. Rossi\footnote{Corresponding author} \\
Istituto Nazionale Fisica Nucleare\\
Sezione di Napoli \\
Naples, Italy\\
biagio.rossi@na.infn.it\\
   \And
F. Antoniali, M. L. Conza, N. Formicola \\
  ANTUR Ricerca e Sviluppo s.r.l.\\
  Via Ferrovia 70, Cercola (Italy) \\
  Cercola, Italy\\
}

\begin{document}
\maketitle

\begin{abstract}
In this paper, a statistical analysis of the performance of a low insulin index, alkaline and functional diet developed by ANTUR evaluated with machine learning techniques is reported. The sample of patients was checked on a regular basis with a BioImpedenziometric (BIA) analysis. The BIA gives about 40 parameters output describing the health  status of the patient. The sample of 1626 patients was grouped in clusters with similar characteristics by a neural network algorithm. A study of the behaviour of the BIA parameters evolution over time with respect to the first visit was performed. More than the 75\% of the patient that followed the ANTUR diet showed an improvement of the health status.
\end{abstract}

% keywords can be removed

\section{Introduction}
In recent decades, according to World Health Organization, the incidence of overweight and obesity in the population has been one of the most serious public health problems in Western countries. Obesity is one of the major risk factors underlying of the premature onset of cardiovascular and cerebrovascular diseases, diabetes, osteoporosis, gastrointestinal diseases and some cancers \cite{obesity_who}.

%\section{Basic principles of the ANTUR diet}
%%%%%%%%%%%%%%%%%%%%%%%%%%%%%%%%%%%%%%%%%%%%%%%%%%%%%%%%%%%
\section{Antur Diet: a low insulin index, alkaline and functional diet}
%%%%%%%%%%%%%%%%%%%%%%%%%%%%%%%%%%%%%%%%%%%%%%%%%%%%%%%%%%%%

The main goal of the Antur diet is to improve patient well-being and reduce body weight by primarily targeting fat mass. Patients were given a low-insulin, alkaline, and functional diet, which was proposed as a useful tool in the prevention of silent inflammation and chronic diseases, as well as in the management of established diseases. The Antur diet aims to be a low-insulin diet, as hyperinsulinemia is associated with overweight~\cite{FAGHFOORI2017S429}, alterations in glucose metabolism, poor lipid profile~\cite{Holt1997-me,NIMPTSCH2011182}, hypertension~\cite{10.1172/JCI111776}, increased risk of early-onset prostate cancer~\cite{Fu2021-pa}, and insulin resistance, a condition typical of various diseases, such as type 2 diabetes mellitus~\cite{6} and polycystic ovary syndrome~\cite{FAGHFOORI2017S429}. In order to ensure the proper functioning of physiological processes, blood pH should be between 7.35 and 7.45~\cite{Quade2021-rr}. Values below 7.35 lead to acidosis, which predisposes to various disorders, including vasodilation, insulin resistance, reduction in neuronal excitability, impairment of the immune system~\cite{Quade2021-rr}, and bone demineralization~\cite{Carnauba2017-rk,9}. The condition of acidosis is due to an unbalanced diet, renal diseases, diabetes, medication consumption, altered renal acid excretion, mineral loss, altered lactate metabolism, and tumors~\cite{Pillai2019-ej, MSD}. The alkaline diet is based on the intake of alkaline foods, negative potential renal acid load (pral) index, that counteract the acidifying effect of positive pral index foods, in order to have an overall alkaline balance and counteract any decreases in pH. The Antur diet is based on the use of foods, considering not only their calories and macro-nutrient content, but also their micro-nutrient content, as these exert (positive or negative) effects in various body areas. The cooking method is also fundamental~\cite{12}, as it must be able to preserve the benefits of the foods without sacrificing taste. Functional nutrition can be a valuable support in the management of various diseases, such as type 2 diabetes mellitus~\cite{Alkhatib2017-uc}, and in limiting the impact of the side effects associated with drug intake~\cite{14}. Fundamental components of the functional diet are spices and herbs for their antioxidant, digestive, lipid- lowering, antibacterial, anti-inflammatory, antiviral, and anticancer properties~\cite{Viuda-Martos2011-vl}. 

Patients followed nutritional plans with the above-mentioned characteristics, but personalized according to their own needs, preferences, the possible presence of diseases and/or food allergies/intolerances, and based on their body weight. Each nutritional plan involved the alternation of {\bf three phases} cyclically repeated until the goal was reached. 
\begin{itemize}
    \item Phase 1 - non-alkaline attack phase in which the patient consumed:
    \begin{itemize}
        \item 1.3 g/kg of body weight of protein
        \item 0.7 g/kg of body weight of carbohydrates
        \item 1.2 g/kg of body weight of lipids
    \end{itemize}
    \item Phase 2 - functional and low-insulin index phase in which the patient consumed:
    \begin{itemize}
        \item 1.2 g/kg of body weight of protein
        \item 1.5 g/kg of body weight of carbohydrates
        \item 1.0 g/kg of body weight of lipids
    \end{itemize}
    \item Phase 3 - alkaline and functional phase in which the patient consumed:
     \begin{itemize}
        \item 1.0 g/kg of body weight of protein
        \item 1.3 g/kg of body weight of carbohydrates
        \item 0.9 g/kg of body weight of lipids
    \end{itemize}
 \end{itemize}  
%
%%%%%%%%%%%%%%%%%%%%%%%%%%%%%%%%%%%%%%%%%
\section{Data taking with the BIA machine} 
%%%%%%%%%%%%%%%%%%%%%%%%%%%%%%%%%%%%%%%%%%

Body weight divided by the square of the height results in Body Mass Index (BMI), which lead to establish whether an individual is underweight, normal weight, overweight or obese. 
However, BMI does not describe body composition neither hydration status. 
Bioimpedance analysis (BIA) is a doubly indirect method to assess body composition. BIA is a safe, non-invasive and cost-effective test; it provides quick and reproducible results and the equipment to perform the exam is easily transported. 
Therefore, BIA is also used in hospital to monitor the nutritional and hydration status of hospitalized patients. 
In this method, body consists in an electrical circuit in which two or more resistors and capacitors are connected in parallel \cite{Kyle2004-qj}. 
BIA allows to quantify total body water (TBW) by measuring body impedance. The interest in the application of bioelectric impedance to human body has developed from the first studies on the relationship between the bioelectric impedance and the content of body water and physiological variables in human and animal tissues. Impedance ($\hat{Z}$) is the frequency-dependent opposition of a conductor to the flow of an alternating electric current. It is a complex quantity, since $\hat{Z} \in \mathcal{C}$, and thus can be defined by its module $Z$ and phase $\phi$.
The impedance module can be derived from:
\begin{equation}
    Z = \sqrt{R^2+X_C^2},
\end{equation}
where $R$ is the resistance and $X_C$ is the reactance \cite{Chumlea1994-lg}.
Body resistance is determined by: 
\begin{equation}
    R = \rho \frac{L}{A},
\end{equation}
where $L$ is the shape/length and $A$ is the surface, and $\rho$ is the intrinsic resistivity area of the body.
In biological systems, resistance is due to the presence of water.
The reactance is the body property to resist voltage variations; it is inversely proportional to the signal frequency ($\omega$) and to the capacitance (C): 
\begin{equation}
    X_C = \frac{1}{\omega C}
\end{equation}
%Capacitance is the property of a circuit to store electric charge [3]. 
%It is the ratio between the amount of electrical charge that may be stored by the capacitor and the voltage \cite{Khalil2014-kp}.

In biological systems the reactance is due to the presence of cell membranes that work as capacitors consisting of two plates separated by an insulating layer that stores electric charges. 
The bioelectric impedance is more influenced by the resistance, because of the greater presence of water than membranes in human body.
The phase angle can be derived from resistance and reactance, as:
\begin{equation}
    \phi = \arctan \left(\frac{X_C}{R}\right) \frac{180^\circ}{\pi}
\end{equation}
The phase angle represents the angle between the impedance vector and the resistance. 
The phase angle is directly proportional to the reactance; hence it is considered an
indicator of cell integrity. 
This parameter is also widely used in hospital to predict the long-term survival of patients suffering from diseases like liver cirrhosis \cite{Selberg2002-tf}, pulmonary diseases \cite{Faisy2000-xn}, HIV infections \cite{Schwenk2000-xg} and AIDS \cite{Ott1995-ad}, several types of cancer \cite{Gupta2004-tn,Toso2000-bs}, dialysis \cite{Maggiore1996-mp,Pupim1999-nk}, bacteremia \cite{Schwenk1998-jv} and sepsis \cite{mattar1995total}. 
The phase angle varies depending on age, sex and ethnicity \cite{Bosy-Westphal2006-dh}. 
This parameter increases as free-fat mass increases \cite{Cancello2023-rf} and it decreases in inflammatory status, because cell membranes can be damaged when oxidative stress occurs \cite{Da_Silva2023-cw} and in elderly subjects \cite{Campa2023-dg}. 
At low frequencies, electric current can only flow through body fluids, while at high frequencies it can penetrate through cell membranes.
Due to wide amount of water and electrolytes, in the Free-Fat Mass (FFM) the observed impedance is slightly lower than measured impedance in Fat Mass \cite{Lukaski1996-yi}.
Originally, BIA assessed body composition by measuring the impedance using a single 50 kHz frequency, generally detected between the wrist and the ankle, as suggested by Nyboer et al. his studies established that 50 kHz was the critical frequency of muscle tissue, hence the current frequency at which the reactance maximum is recorded \cite{Nyboer1970}.
According to the axioms of impedance plethysmography, the square of the height (the “length” of the “conductor”) divided by the total body resistance results in impedance index, which is a total conductive volume index. This index can describe the FFM volume, due to higher electrolyte content and higher conductivity of FFM compared to adipose tissue \cite{Kay1956-bi}. Impedance is mainly determined by resistance, which depends on the amount of tissue fluids; therefore the single-frequency bia (Sf-BIA) is able to determine the total body water and the lean mass that is 73\% hydrated. 
Fat mass is calculated by the difference between body weight and lean mass \cite{Guo1987-gr}. In the 1960s Thomasset et al. established that it is possible to discriminate between intracellular and extracellular water by using different frequencies (both high and low frequencies); this allows a more accurate estimation of body composition \cite{Thomasset1962-xd}.
At first human body was considered as a single cylindrical block with uniform cross-sectional area, but a better approximation describes the human body as a sum of five different cylinders (two upper limbs, two lower limbs and the trunk). In order to clarify how every body segment affects the total body impedance, Segmental BIA method was developed \cite{Lorenzo2003-cw}.
To estimate body composition, InBody 570 was used in this study. It uses the Direct Segmental Multifrequency Bioimpedance Analysis (DSM-BIA), an accurate and precise technology which allows a multi-frequency analysis for each body segment separately; therefore, impedance values are provided for each body district and for each frequency used. 
InBody 570 is equipped of eight tactile and fixed electrodes, allowing the electrical circuits to always maintain the same length which results in a high level of accuracy and reproducibility. In addition, InBody570 evaluate data derived from the measured impedance, avoiding empirical estimates; age and sex only define confidence intervals for each parameter \cite{Thomas2001}.

\section{Statistical description of the sample of patients}
\label{sec:sampledescription}
In this section a statistical description of the patients samples is given. The variables recorded for each patient and for each BIA test are 23 and they are categorised in 6 main groups: Hydration, Circumferences, Adiposity, Body Composition, Muscularity e Metabolism. 
As can be seen in Table \ref{tab:BIA_variabili}, Hydration group contains ECW (Extra Cellular Water), TBW (Total Body Water) and ICW (Intra-Cellular Water) variables, the Circumferences group contains the dimension (in cm) of waist, neck, chest, hips, arm, thigh, the body composition group is formed by the BMI (Body Mass Index) and the WHR (Waist-to-Hip ratio), the Muscularity group is composed by SMI (Skeletal Muscle Index) and SMM (Skeletal Muscle Mass), the Adiposity group is composed by FFM, Visceral Fat Level (VFL), Body Fat Mass (BFM), Obesity degree variables, and Metabolic group which is formed by Body Cell Mass (BCM), Minerals, Protein, Basal Metabolic Rate (BMR).
%
%\begin{figure}[!htb]
%    \centering  
%    \includegraphics[scale=0.5]{pics/BIA_variabili.png}
%    \caption{Sketch of the 6 groups of variables obtained from each BIA test.}
%    \label{fig:BIA_variabili}
%\end{figure}

\begin{table}[!htb]

\definecolor{Green}{rgb}{0.82,0.91,0.68}
\definecolor{Purple}{rgb}{0.85,0.8,90}
\definecolor{Orange}{rgb}{0.94,0.74,0.44}
\definecolor{Pink}{rgb}{0.91,0.75,0.75}
\definecolor{Gray}{rgb}{0.89,0.93,0.92}
\definecolor{Blue}{rgb}{0.86,0.96,0.96}
\definecolor{lightblue}{rgb}{0.39,0.67,0.88}

\newcolumntype{1}{>{\columncolor{Green}}c}
\newcolumntype{2}{>{\columncolor{Purple}}c}
\newcolumntype{3}{>{\columncolor{Orange}}c}
\newcolumntype{4}{>{\columncolor{Pink}}c}
\newcolumntype{5}{>{\columncolor{Gray}}c}
\newcolumntype{6}{>{\columncolor{Blue}}c}

\renewcommand{\arraystretch}{1.1}
\scalebox{0.67}{
%\begin{tabular}{1 | 2 | 3 | 4 | 5 | 6 }
\begin{tabular}{1 | 2 | 3 | 4 | 5 | 6 }
\hline
\rowcolor{lightblue}
\multicolumn{6}{c}{\textbf{\textcolor{white}{Data}}} \\
\hline
\textbf{Hydration} & \textbf{Adiposity} & \textbf{Metabolism} & \textbf{Muscularity} & \textbf{Body Composition} & \textbf{Circumferences} \\
\hline
 %& & & & & & & & & &\\
Extra-Cellular Water (ECW) & Fat Free Mass (FFM) & Body Cell Mass (BCM) & Skeletal Muscle Index (SMI) & Body Mass Index (BMI) & Waist \\
Total Body Water (TBW) & Visceral Fat Level (VFL) & Minerals & Skeletal Muscle Mass (SMM) & Waist-to-Hip Ratio (WHR) & Neck \\
Intra-Cellular Water (ICW) & Body Fat Mass (BFM) & Protein &  & & Chest \\
 & Obesity Degree & Basal Metabolic Rate (BMR) & & & Hips \\
  & & & & & Arm \\
  & & & & & Thigh \\
 %  & & & & & & & & & &\\
  \hline
  \end{tabular}}
\\
\caption{Summary table of the 6 groups of variables obtained from each BIA test.}
  \label{tab:BIA_variabili}

\end{table}

Each patient after the first BIA test receives a prescription treatment protocol (ANTUR diet) to be followed in order to improve its health status. The protocol contains a diet, advice for physical activity and, more generally, for a healthier lifestyle.
The data analysis presented here has the goal of studying the behaviour of the BIA variables, connected to the health status and the wellness of the patient, as a function of the time from the first BIA test. 

The data samples of 9080 tests undergo a first cleaning by excluding the records with: 
\begin{itemize}
    \item a single BIA test; 
    \item broken and/or incomplete registration data.
\end{itemize}
After the first cleaning-up, the data sample is composed by 1626 patients of which 424 are male and 1202 are female, as reported in Table~\ref{tab:pazienti}. The total number of records (BIA tests performed on the patients' sample) in 6 months is 6664. 
\begin{table}[!ht]
  \centering
  \begin{tabular}{lccc}
    \toprule
         & Male & Female & Total \\
    \midrule
    Patients &  424  & 1202 & 1626 \\
    Total BIA tests & 1436 & 5520 & 6664 \\
    \bottomrule
  \end{tabular}
  \\
  \caption{Data sample information: analysis records divided by male and female.}
  \label{tab:pazienti}
\end{table}

\subsection{Characterization of the data samples} 
\begin{figure}[htb]
    \centering  
    \includegraphics[width=0.8 \textwidth]{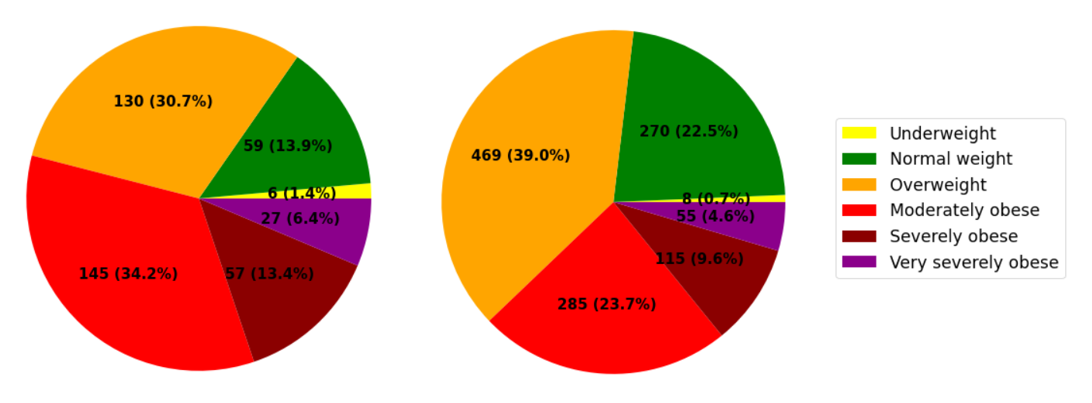}
    \caption{BMI distribution of the sample at the first BIA test (trial 0) for males (on the left) and for females (on the right).}
    \label{fig:BMI_generale}
\end{figure}

Taking as reference values' for the BMI the Tab.~\ref{tab:BMI}, the full sample of patients (1626 people), split in male and female and before the nutrition treatment start (trial 0), have the BMI distribution showed in pie charts in Figure~\ref{fig:BMI_generale}. 
\begin{table}[htb]
    \centering
    \begin{tabular}{cc}
 \toprule
Status & BMI \\
 \midrule
Underweight        & <18.5 \\
Normal weight       & 18.6-24.9 \\
Overweight         & 25.0-29.9 \\
Moderately obese   & 30.0-34.9 \\
Severely obese   & 35.0-39.9 \\
Very severely obese   & >40 \\
\bottomrule
    \end{tabular}
    \\
    \caption{BMI reference value.}
    \label{tab:BMI}
\end{table}
About 15\% of males (23\% of females) are underweight/normal weight. They need a maintenance diet for the weight and to improve other biologic parameters for an improved wellness. The 31\% (39\%) are overweight, while the remaining 54\% (38\%) shows obesity of some level. This represents the starting point dataset on which the ANTUR nutrition treatment has been applied.

The BMI is a good indicator but it's not the only relevant one. Another important parameter is the waist-to-hip ratio (WHR), commonly used as a marker of cardiovascular risk. Reference values for WHR are reported in Tab.~\ref{tab:WHR}. 

\begin{figure}[htb]
    \centering  
    \includegraphics[width=0.8 \textwidth]{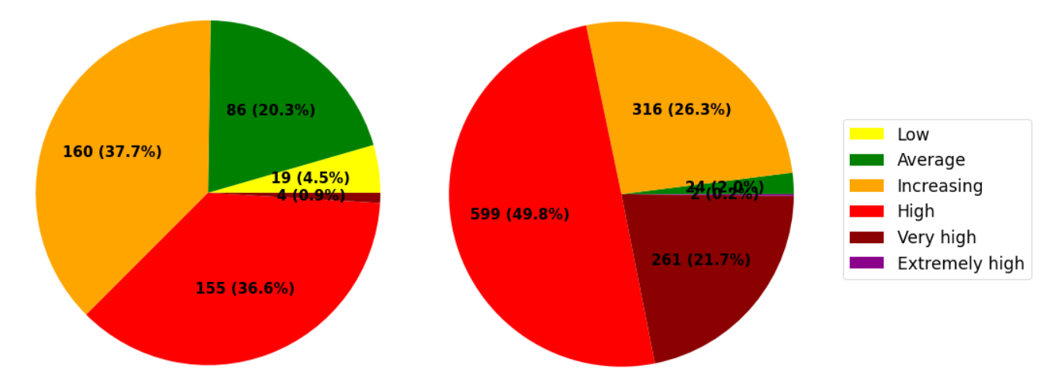}
    \caption{Weist-to-Hip ratio (WHR) sketch of the sample at the first BIA test (trial 0) for males (on the left) and for females (on the right).}
    \label{fig:WHR_generale}
\end{figure}

\begin{table}[htb]
    \centering
    \begin{tabular}{ccc}
    \toprule
    Risk level & Male & Female \\
    \midrule
    Low            & $<0.80$     & $<0.70$\\
    Average    & 0.81-0.90 & 0.71-0.80\\
    Increasing     & 0.91-0.99 & 0.81-0.89\\
    High           & 1.00-1.19 & 0.90-1.09\\
    Very high      & 1.20-1.29 & 1.10-1.19\\
    Extremely high & $>1.30$     & $>1.20$\\
    \bottomrule
    \end{tabular}
    \\
    \caption{WHR reference value.}
    \label{tab:WHR}
\end{table}
Before starting the ANTUR treatment, about 25\% of males (and only 2\% of females) shows low/average risk with respect to WHR thresholds. About 38\% (26\%) have an increasing risk, while the 37\% (72\%) shows a high/very high/extremely high risk, as reported in the pie charts in Figure~\ref{fig:WHR_generale}.

Moreover to have a more detailed overview of the used dataset mean, standard deviation and percent error values for all the variables recorded at the first BIA test (trial 0) for all the
patients are reported in table \ref{tab:first_visit_meanstd_all}.
\begin{table}[!htb]
    \centering
    \begin{tabular}{lccc}
    \toprule
    Variable:  & Mean & Standard deviation & Percent error [\%]\\
    \midrule
    %Mean values to the first visit:  & 0 \\
    Circumference arm [cm] & 34.81  & 5.46  & 16 \\
    Circumference neck [cm] & 37.88  & 4.24  & 11 \\
    Circumference chest [cm] & 100.74  & 9.94  & 10 \\
    Circumference waist [cm] & 98.25  & 15.56  & 16 \\
    Circumference hip [cm] & 103.37  & 8.85  & 9 \\
    Circumference right thigh [cm] & 55.98  & 5.54  & 10 \\
    Gender [fraction of females] & 0.74  & 0.44  & 59 \\
    Height [cm] & 164.08  & 8.87  & 5 \\
    Age & 38.38  & 14.98  & 39 \\
    Weight [kg] & 79.76  & 18.59  & 23 \\
    ICW [\%] & 22.62  & 4.77  & 21 \\
    Proteins & 9.78  & 2.06  & 21 \\
    Minerals & 3.45  & 0.72  & 21 \\
    SLM & 46.87  & 9.78  & 21 \\
    BMI [kg/m$^2$] & 29.49  & 5.75  & 20 \\
    PBF & 36.71  & 8.87  & 24 \\
    FFM right arm & 2.72  & 0.77  & 28 \\
    FFM left arm & 2.69  & 0.76  & 28 \\
    FFM trunk & 22.58  & 4.65  & 21 \\
    FFM right leg & 7.51  & 1.72  & 23 \\
    FFM left leg & 7.48  & 1.68  & 22 \\
    BFM right arm & 2.52  & 1.73  & 69 \\
    BFM left arm & 2.53  & 1.74  & 69 \\
    BFM tronco & 15.03  & 5.60  & 37 \\
    BFM right leg & 4.28  & 1.61  & 38 \\
    BFM left leg & 4.26  & 1.59  & 37 \\
    Target weight [kg] & 63.01  & 10.62  & 17 \\
    BMR & 1444.03  & 223.61  & 15 \\
    WHR & 0.95  & 0.08  & 9 \\
    VFL & 13.27  & 5.06  & 38 \\
    Obesity degree & 139.63  & 26.87  & 19 \\
    AMC & 29.06  & 4.07  & 14 \\
    BMC & 2.85  & 0.60  & 21 \\
    Circumference right arm [cm] & 34.89  & 5.46  & 16 \\
    Circumference left arm [cm] & 34.81  & 5.46  & 16 \\
    Circumference left thigh [cm] & 55.64  & 5.26  & 9 \\
    SMI & 7.48  & 1.10  & 15 \\
    Recommended calories [kcal] & 2008.34  & 498.46  & 25 \\
%    Cluster & 5.72  & 2.00  & 35 \\
    BFM [\%] & 36.70  & 8.88  & 24 \\
    FFM [\%] & 63.30  & 8.88  & 14 \\
    SMM & 34.91  & 5.12  & 15 \\
    BCM & 41.24  & 5.90  & 14 \\
    TBW & 46.45  & 6.43  & 14 \\
    ECW [\%] & 61.45  & 1.95  & 3 \\
    \bottomrule
    Total number of patients  & 1626 \\
    \end{tabular}
    \\
    \caption{Mean, standard deviation and percent error values for all the variables recorded at the first BIA test for all the patients.}
    \label{tab:first_visit_meanstd_all}
\end{table}
%
%%%%%%%%%%%%%%%%%%%%%%%%%%%%%%%%%%%%%
\section{Machine learning to obtain clusters of patients}
\label{sec:SOMnet}
%%%%%%%%%%%%%%%%%%%%%%%%%%%%%%%%%%%%%

\begin{figure}[!htb]
    \centering  
    \includegraphics[width=.7 \textwidth]{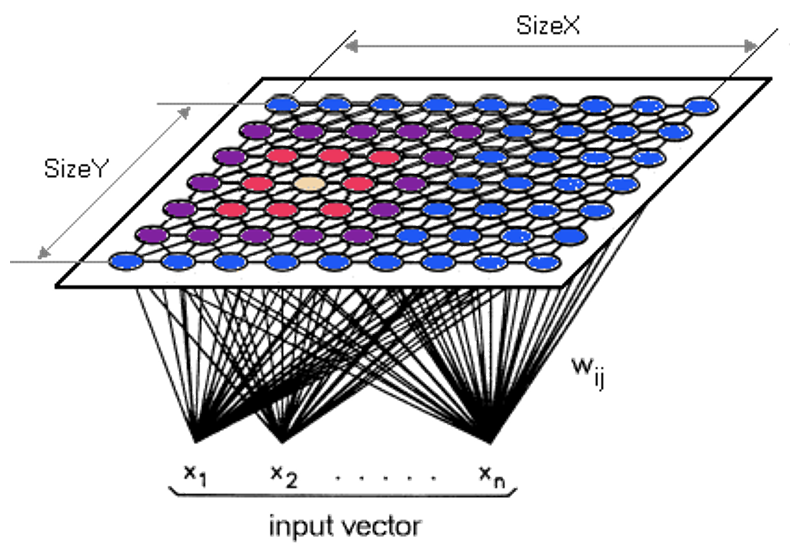}
    \caption{Schematic view of a SOM architecture. In most cases, n $\times$ the number of input variables is much bigger than the 2 dimensions associated to the output grid.}
    \label{fig:SOM_scheme}
\end{figure}

To evaluate the performance of the ANTUR approach a Self-organizing map has been developed to cluster patient with common characteristics. Self-organizing maps (SOMs) are a special type of artificial neural networks trained using unsupervised learning. They allow the data sample to be divided into classes with similar properties. 
The data sample is branched into a training, validation and test samples. The training sample is a subset of tge dataset used for the learning phase. Validation is a technique in machine learning to evaluate the performance of models during learning and the validation sample is a subset of examples used to tune the parameters of the network. The test sample is a subset of data not used in the learning phase used to evaluate the results of the developed network.
The SOM network makes it possible to produce a representation of the data sample provided as input, generally high-dimensional, in a low-dimensional space while preserving its topological properties. This property makes SOMs particularly useful for visualizing high-dimensional data.
The model was first described by Teuvo Kohonen, and it is often referred to it as Kohonen maps~\cite{Kohonen1982}.
Self-organizing maps are laterally connected neural networks where output neurons are organized in low-dimensional (usually 2D or 3D) grids. Each input is connected to all output neurons, such that each output neuron j is associated with a weight vector $W_j$ of the same size as the inputs vector, as can be seen from Figure \ref{fig:SOM_scheme} where a schematic view of the network architecture is shown. The size of the input vector is generally high and the output used in this work is a 2D map.

The goal of learning in SOMs is to specialize different parts of the grid to respond similarly to particular input patterns. This is inspired in part by how visual, auditory, or other sensory information are handled by separate parts of the cerebral cortex in the human brain.
SOMs learn to classify the data provided as input to the network based on the identification of common features obtained by analyzing all the variables provided. 
Therefore, SOMs learn both distribution (as well as competitive levels) and the topology of the input vectors on which they are trained. The SOM networks are used for various applications, of which, one of the main ones, is clustering that is, the partitioning of a dataset into a collection of classes, or clusters, with similar characteristics.
The way these networks train is through a competitive mechanism, where all of the output neurons take part in a "winner takes all" competition. The winner neuron j is the one having the maximum output value Y, defined as

\begin{equation}
    Y_j = \sum_{i}w_{ij}*x_i
\end{equation}

where $w_{ij}$ is the weight matrix of the network and $x_i$ is the inputs vector. This analytic relation shows that the chosen neuron is also the one for which the weight vector $W_j$ is the most similar to $x_i$. Once the winner is established, the weight vectors are updated only for this neuron and for those belonging to its neighbourhood (active neurons), defined by an ad hoc neighbouring function H. The difference depends on the distance from the winning neuron, so that weights associated to farthest neurons are changed less than the closest ones \cite{VESANTO1999111}. Through this mechanism, groups of neurons with similar properties are created. The training process is interrupted when the SOM output stops changing or no significant changes are observed.

This type of network is particularly indicated when working with a not homogeneous sample, like the one described in Sec.~\ref{sec:sampledescription} and used in this work. 
A SOM network is applied to the dataset of the patients by dividing the sample in training, and test with a ratio of 66:34.
The parameters of the network applied are:
\begin{itemize}
    \item $5\times4$ grid topology: 20 neurons
    \item Euclidean distance metric
    \item weights modification occurs within a maximum range of three nodes
    \item 5000 training epochs
    \item 18 input variables.
\end{itemize}

\begin{figure}[!htb]
    \centering  
    \includegraphics[width=.48 \textwidth]{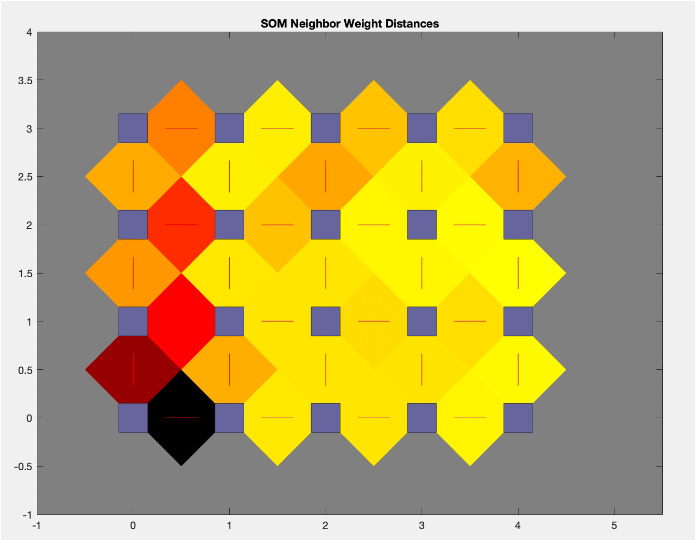}
    \includegraphics[width=.48 \textwidth]{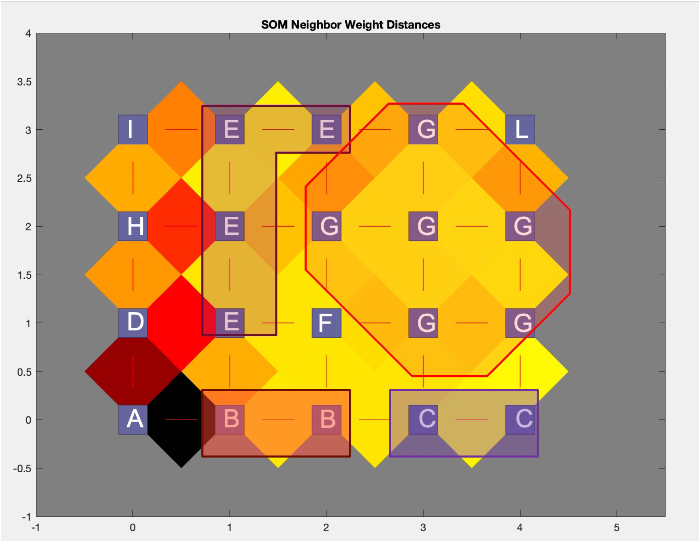}
    \caption{Output of the SOM network (on the left) and sketch of the definition of the cluster (on the right). Clusters here are labelled with letters from A to L, while in the following sections they will be labelled with numbers from 1 to 10.}
    \label{fig:SOM_def}
\end{figure}

\begin{figure}[!htb]
    \centering  
    \includegraphics[width=.45 \textwidth]{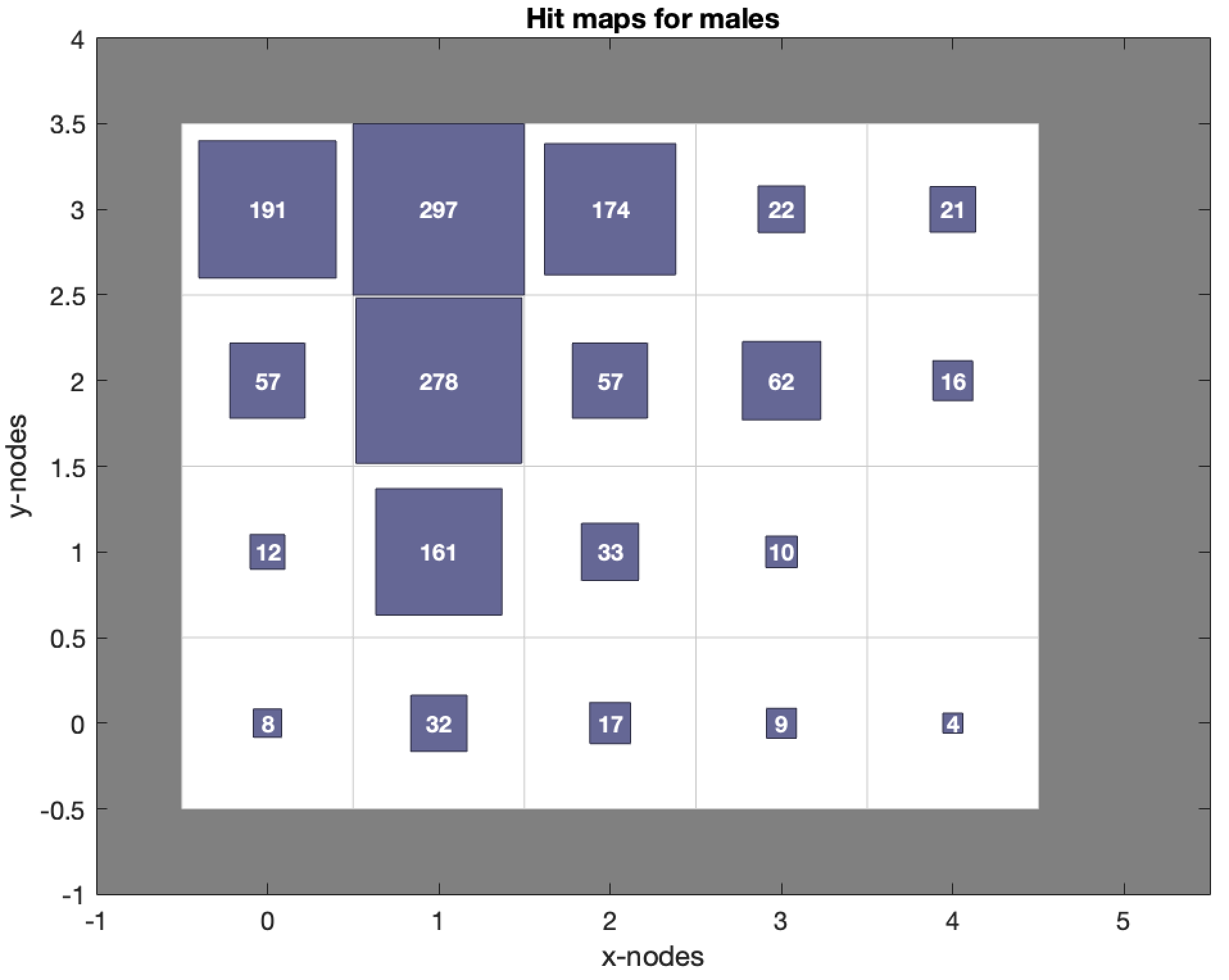}
    \includegraphics[width=.45 \textwidth]{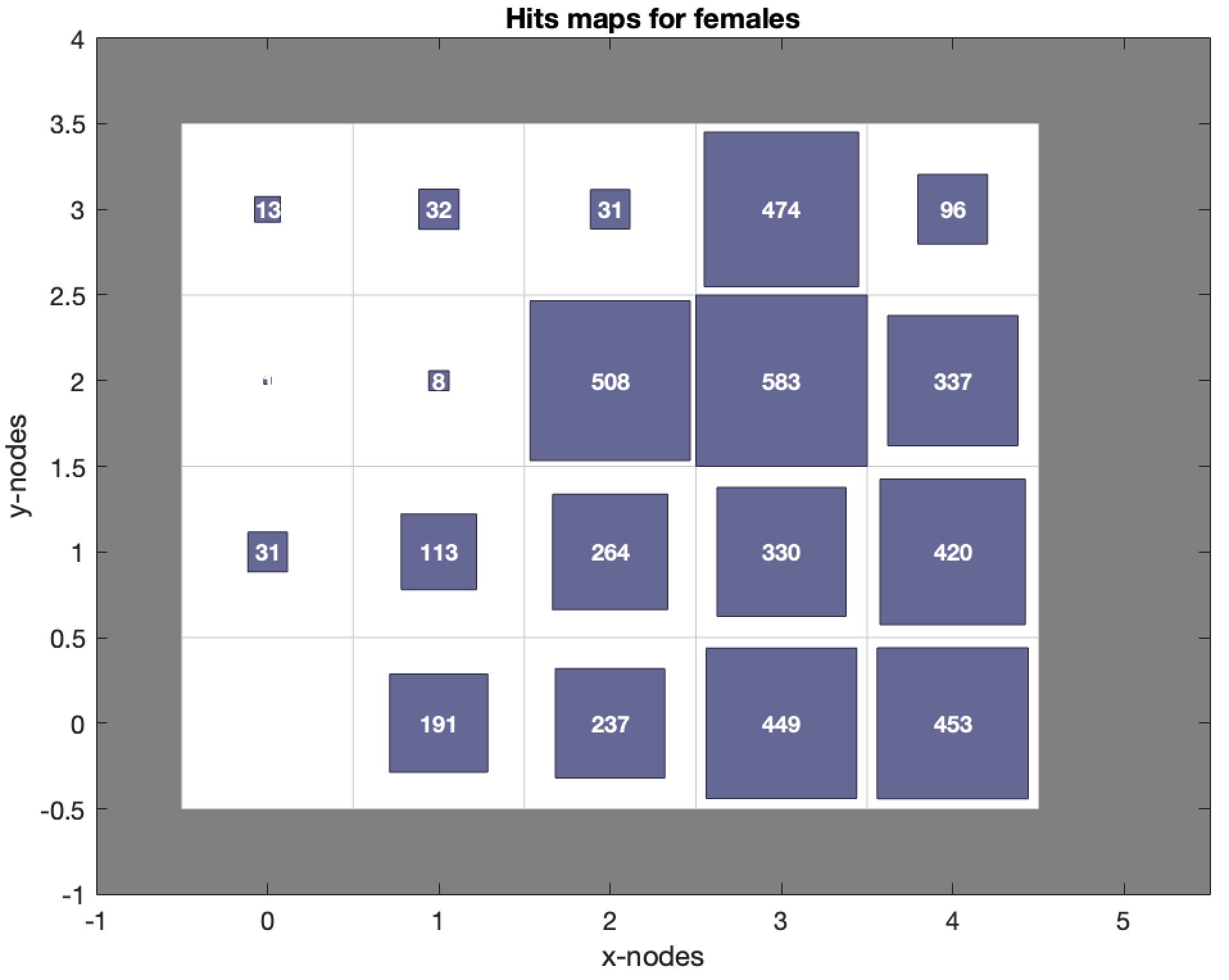}
    \includegraphics[width=.45 \textwidth]{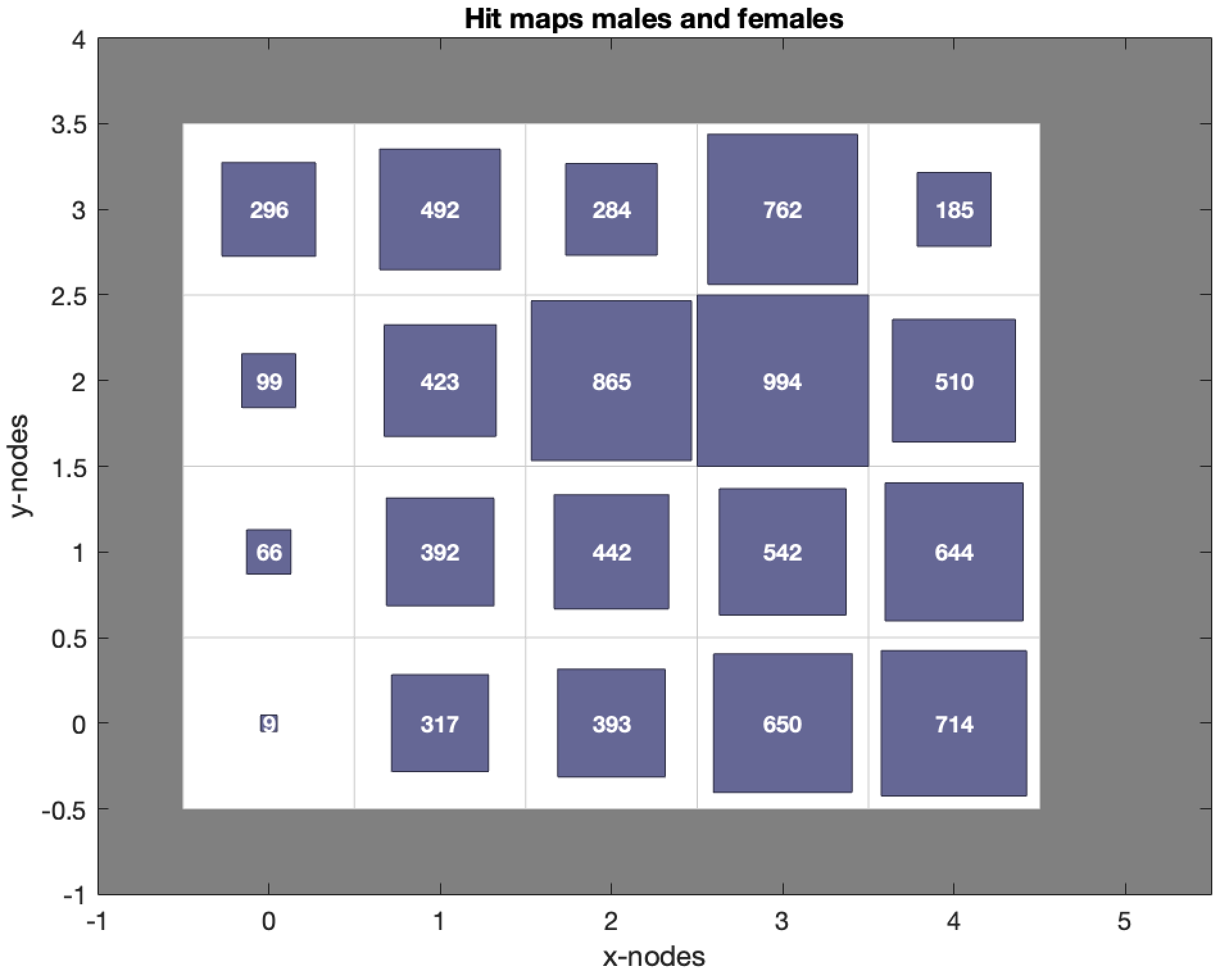}
    \includegraphics[width=.45 \textwidth]{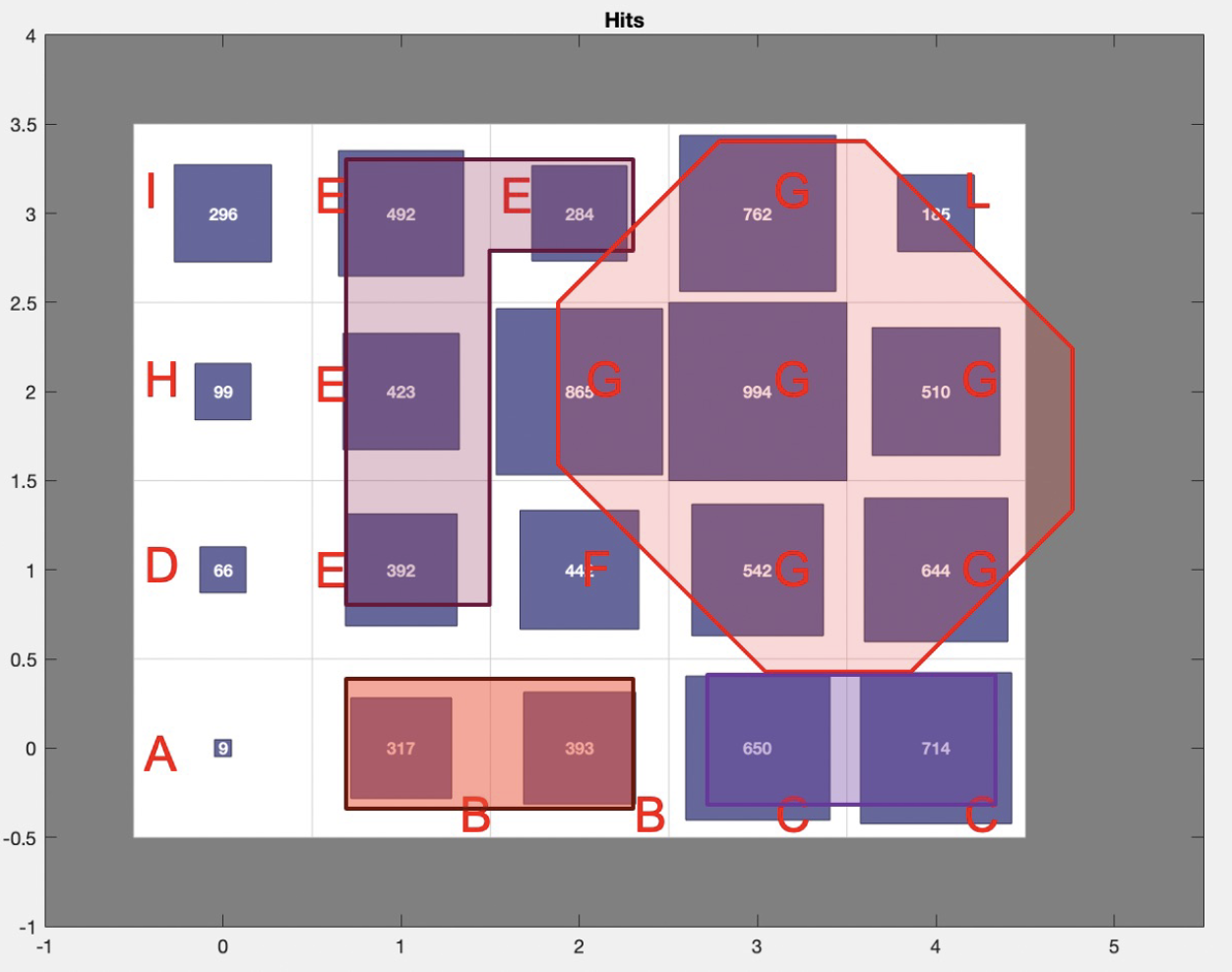}
    \caption{Hit maps for males (upper left) and females (upper right) in the training dataset. The lower hit maps are for the total number of patients (left) and the same with the superimposed cluster definition (right).}
    \label{fig:SOM_hit}
\end{figure}

Figure~\ref{fig:SOM_def} shows the grid topology of the SOM network used (on the left) and the cluster definition performed after the training (on the right).
In this figure, the squares represent the output neurons connected by red lines, while the color indicate the distance between them: a lighter color (yellow) is associated with a short distance, whether a darker color (red-black) is associated with a larger distance. Here, the distance is the euclidean distance in the $x-y$ plane. By looking for regions in the grid with yellow color enclosed in red colored segment strips, 10 clusters were identified labelled with letters from A to L. In the following paragraphs, they will be referred to with numbers from 1 to 10.
\\
Moreover, in Figure \ref{fig:SOM_hit}, the hit maps for males, females and both kinds of patients are reported. In these, again, each square is an output neuron, and the numbers represent how many data features among the totality of network inputs are associated to each neuron. They show that the clusters not only present a remarkable difference between each other in their unique characteristics, but we can also see that some clusters, hence some characteristics, are mostly associated to males (upper-left region) while others are mostly associated to females (lower-right region). %The dimension ratio between these two regions is due to the number ratio between males and females patients.

\section{Statistical description of the ML clusters}
A total of 10 clusters are defined according to the SOM network defined in Sec.~\ref{sec:SOMnet}.
The cluster defined are characterized by analyzing the features used to train the SOM network and Table~\ref{tab:cluster_mean_val} reports the features mean values for the first BIA test. 

\begin{table}[!htb]
\resizebox{\textwidth}{!}{
\begin{tabular}{l c c c c c c c c c c}
%Valore medio alla prima visita per il cluster:  & 1 & 2 & 3 & 4 & 5 & 6 & 7 & 8 & 9 & 10 \\
Mean value for the  cluster:  & 1 & 2 & 3 & 4 & 5 & 6 & 7 & 8 & 9 & 10 \\
\toprule
Gender [fraction of females] & 0.00  & 0.89  & 0.99  & 0.62  & 0.21  & 0.90  & 0.94  & 0.04  & 0.08  & 0.65 \\
Age & 41.50  & 44.39  & 55.00  & 39.16  & 38.52  & 27.11  & 34.57  & 38.09  & 36.06  & 22.90 \\
Height [cm] & 182.00  & 160.98  & 157.63  & 164.00  & 173.83  & 161.23  & 161.49  & 179.12  & 178.45  & 152.26 \\
Weight [kg] & 184.50  & 100.60  & 76.14  & 129.14  & 86.60  & 87.31  & 67.96  & 133.15  & 109.92  & 43.54 \\
BMI [kg/m$^2$] & 55.80  & 38.83  & 30.66  & 48.07  & 28.72  & 33.62  & 26.04  & 41.45  & 34.56  & 18.82 \\
BFM [\%] & 50.30  & 49.15  & 43.38  & 52.20  & 28.63  & 45.51  & 34.44  & 42.26  & 36.76  & 20.54 \\
FFM [\%] & 49.70  & 50.85  & 56.62  & 47.80  & 71.37  & 54.49  & 65.56  & 57.74  & 63.24  & 79.46 \\
%Arm circumference  [cm] & 83.25  & 42.26  & 34.90  & 52.06  & 35.20  & 37.52  & 31.69  & 46.51  & 39.88  & 24.64 \\
Neck circumference  [cm] & 50.30  & 44.47  & 37.96  & 50.59  & 38.93  & 40.27  & 35.03  & 46.77  & 42.69  & 29.24 \\
Chest circumference [cm] & 130.30  & 113.87  & 100.04  & 126.27  & 104.99  & 105.87  & 93.92  & 123.75  & 115.22  & 78.22 \\
Waist circumference [cm] & 153.53  & 119.84  & 100.17  & 134.27  & 99.17  & 107.73  & 88.40  & 137.94  & 120.62  & 66.56 \\
Hip circumference [cm] & 141.35  & 115.71  & 102.94  & 128.84  & 104.86  & 109.35  & 97.86  & 123.68  & 114.55  & 84.54 \\
Right arm circumference [cm] & 82.88  & 42.41  & 34.99  & 52.13  & 35.29  & 37.58  & 31.74  & 46.70  & 39.89  & 24.72 \\
Left arm circumference [cm] & 83.25  & 42.26  & 34.90  & 52.06  & 35.20  & 37.52  & 31.69  & 46.51  & 39.88  & 24.64 \\
Right thigh circumference [cm] & 73.23  & 62.12  & 54.78  & 68.74  & 57.45  & 60.05  & 52.99  & 67.18  & 63.09  & 44.62 \\
Left thigh circumference [cm] & 69.95  & 61.56  & 54.55  & 66.93  & 57.12  & 59.49  & 52.77  & 65.87  & 62.53  & 44.60 \\
\bottomrule
Total number of patients  & 4 & 136 & 224 & 16 & 300 & 107 & 707 & 24 & 73 & 34 \\
\end{tabular}}
\\
\caption{Mean values of the most important BIA variables at the first visit for patients belonging to the 10 clusters, as categorized by the SOM network outcome.}
\label{tab:cluster_mean_val}
\end{table}

By referring to Tab.~\ref{tab:cluster_mean_val} one can notice that:
\begin{itemize}
    \item cluster 1 is exclusively characterized by males with a very high BMI;
    \item cluster 2 is largely characterized by medium-aged females with a quite high BMI;
    \item cluster 3 is almost exclusively characterized by medium-aged females with a high BMI;
    \item cluster 4 is largely characterized by males and females with a very high BMI;
    \item cluster 5 is mainly characterized by males with a BMI slightly above average;
    \item cluster 6 is largely characterized by young females with a high BMI;
    \item cluster 7 is largely characterized by females with a BMI slightly above average;
    \item cluster 8 is mainly characterized by males with a medium-high BMI;
    \item cluster 9 is largely characterized by young males with a high BMI;
    \item cluster 10 is mainly characterized by very young males and females with a BMI slightly below average.
\end{itemize}
%All the patients followed the functional low-insulin diet proposed by the ANTUR society but with different targets, according to the belonging cluster. 
%So cluster 1 or cluster 4 should dramatically lower the BMI index by losing weight, regularize the proportion between FFM and BFM, and improve hydration of the body, while cluster 10 should increase BMI and FFM. 
%Even with so different targets the functional low-insulin diet can be tuned in order to fulfill different requirements.
%The objective of this project is to carry out a statistical study of biological parameters from BIA.
%The proposed analysis follows the Knowledge Discovery in Databases (KDD) process for the search for new knowledge from data. 
%The main strategy aims at developing a model able to describe the evolution of the BIA parameters as a function of the time from the beginning of the ANTUR treatment.

All patients adhered to the functional low-insulin diet prescribed by the ANTUR society, but their specific goals varied depending on their assigned cluster. For instance, patients in clusters 1 or 4 aimed to reduce their BMI, regulate the proportion of FFM and BFM, and improve hydration. Conversely, patients in cluster 10 sought to increase their BMI and FFM. Despite the disparate objectives, the functional low-insulin diet can be tailored to meet diverse requirements.

The aim of this project is to conduct a statistical analysis of biological parameters measured through bioelectrical impedance analysis (BIA). The analysis will follow the Knowledge Discovery in Databases (KDD) \cite{KDD} process to uncover new insights from the data. The main strategy is to develop a model that can describe how the BIA parameters change over time since the start of the ANTUR treatment.

The initial step involved reporting biological parameters for each patient as a function of the elapsed time since their first visit. Next, a linear model was introduced to capture the temporal trend, which can be expressed as:
\begin{equation}
    y = a + bx,
\end{equation} 
where $y$ represents the biological variable under investigation, $x$ denotes the time interval from the first visit, $b$ indicates the variation of the biological variable over time, and $a$ represents the initial value of the biological variable.

Since these two parameters are unknown, a least squares procedure is performed to determine their values. This procedure identifies the best-fit straight line that describes the patient's data. The obtained parameters are interpreted as follows:
\begin{itemize}
    \item $a$ is the starting value of the biological variable at the beginning of treatment, as predicted by the statistical inference. 
    \item $b$ represents the daily rate of variation of the BIA parameter, which can be easily translated into the monthly variation rate.
\end{itemize}

%
%%%%%%%%%%%%%%%%%%%%%%%%%%%%%%%%%%%%%%
\subsection{Time behavior of the main BIA parameters}
%%%%%%%%%%%%%%%%%%%%%%%%%%%%%%%%%%%%%%
\label{sec:time_analysis}
This study analyzed the relationship between the biological variables and the elapsed time from the start of the nutritional treatment to understand the variation of biological variables with time and to assess the effectiveness of the dietary protocol.
The main focus was on the following six sectors:
\begin{itemize}
    \item Body composition and cardiovascular disease risk 
    \item Hydration
    \item Adiposity
    \item Muscularity
    \item Circumferences (neck, waist, thigh, etc)
    \item Target weight.
\end{itemize}

\begin{figure}[htb]
    \centering  
    \includegraphics[scale=0.75]{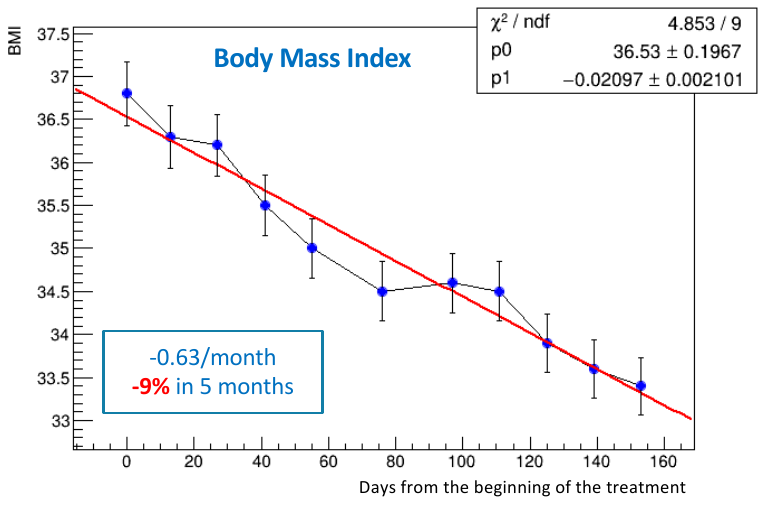}
    \caption{Time dependence of the BMI (kg/m$^2$) for the patient ID1556.}
    \label{fig:BMI_ID1556}
\end{figure}

As example, in Figure~\ref{fig:BMI_ID1556}, the BMI of a male patient with a 2$^{\rm nd}$ degree obesity belonging to the cluster 1. The vertical axis is the BMI (expressed in kg/m$^2$), while on the horizontal axis the elapsed time from the beginning of the nutritional treatment is reported. This patient started with a very high BMI of 36.8 kg/m$^2$ and after 5 months of treatment it decreased to 33.5 (a decrease of 9\%). The plot shows that the BMI decreases linearly with time, which was observed overall for every patient in similar conditions. The data points of BMI values at different medical trials were fitted with a linear model using equation:
\begin{equation}
  \mathrm{BMI_t} = \mathrm{BMI_0} + b \cdot t,  
 \label{eq:BMI_fit}  
\end{equation}
where BMI$_{0}$ is the BMI value at the beginning of the treatment, $t$ is the time elapsed from the first medical trial, BMI$_{t}$ is the BMI value at a given time $t$, while $b$ is the slope of the straight line representing the $\Delta$BMI/day. 
For this patient, the resulting slope value is $b=-0.021$ BMI/days, indicating a consistent improvement of -0.6 BMI points per month over 5 months of treatment. If the trend continues, the patient would reach the status of "Overweight" in 10 months and "Normal weight" in 18 months.

A similar analysis has been performed for the WHR as a function of time. As depicted in Fig.~\ref{fig:BMI_ID1556}, the patient started the treatment ranked as "high risk" (WHR$=1.05$) for cardiovascular disease. After 5 months of treatment, its WHR decreased to 0.97 (8\% reduction), ranked as "Increasing risk". It is clear from the plot that also WHR decreases linearly with time during the treatment. Thus, we performed a linear fit with the equation:
\begin{equation}
  \mathrm{WHR_t} = \mathrm{WHR_0} + c \cdot T,
 \label{eq:WHR_fit}  
\end{equation}
where WHR$_{0}$ is the WHR value at the beginning of the treatment, $T$ is the time elapsed from the first medical trial, WHR$_{t}$ is the WHR value at a given time $t$, while $c$ is the slope of the line and it represents the $\Delta$ WHR/day. In this particular case, the resulting value of the fit slope is $c=-4.6\cdot10^{-4}$ WHR/day. This indicates that the patient got a constant improvement of about -0.014 WHR points per month, over 5 months of treatment. If the trend stays constant one can foresee that the patient in 10 months would be ranked "In Average" while after 18 months he would be ranked "Low risk".   
\begin{figure}[htb]
    \centering  
    \includegraphics[scale=0.75]{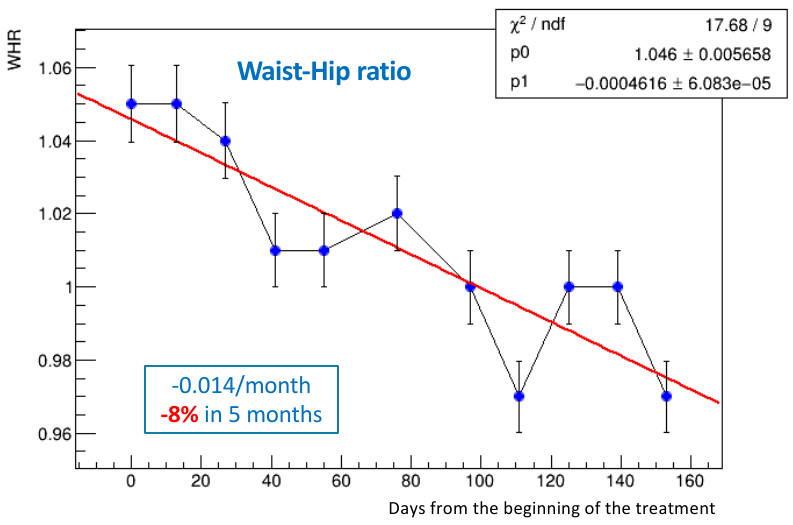}
    \caption{Time dependence of the WHR for the patient ID1556.}
    \label{fig:WHR_ID1556}
\end{figure}

\begin{figure}[htb]
    \centering  
    \includegraphics[scale=0.55]{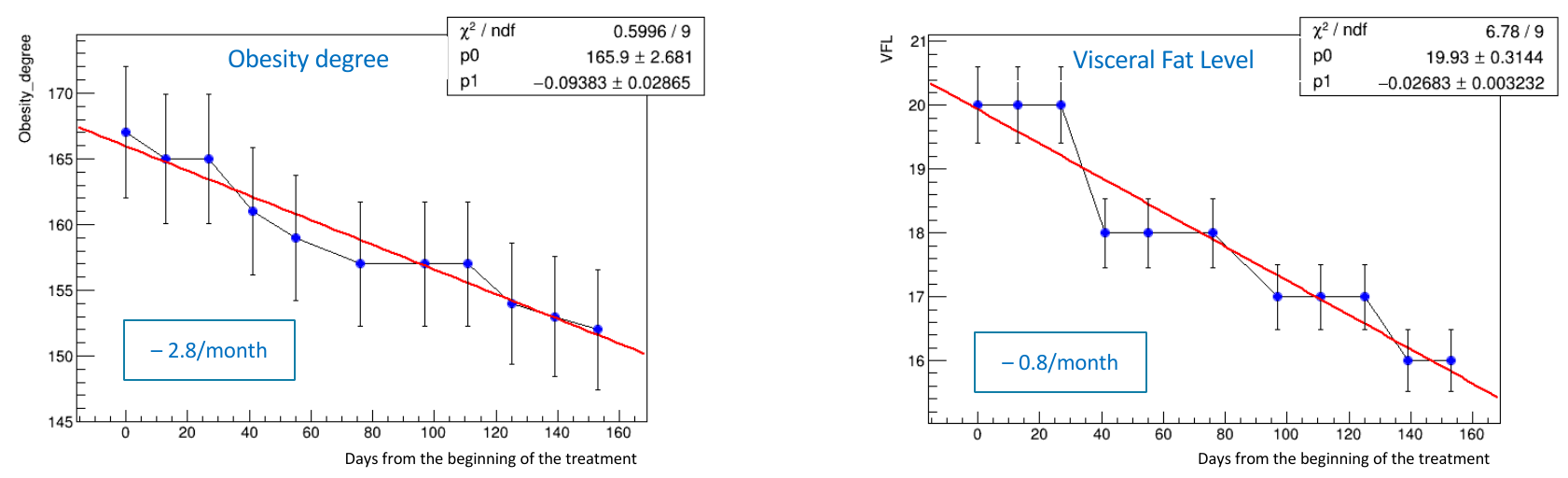}
    \caption{Trend line of the obesity degree, on the left, and  VFL, on the right, as a function of the time from the beginning of the treatment. Both trends are linear with a 10\% reduction of the obesity degree and  a 20\% reduction of the VFL in 5 months of treatment for the patient ID1556.}
    \label{fig:adiposity}
\end{figure}

For what concerns the Body Adiposity, a study of the trends of the obesity degree and of the visceral fat level (VFL) have been performed. In Figure \ref{fig:adiposity} the data points and the trend lines for both obesity degree and VFL are reported. Also in this case, both the variables follow a linear trend with a 10\% reduction for the obesity degree and a 25\% reduction fot the VFL. In particular, the patient with ID 1556 was able to reduce the VFL from 20, classified as "very high" as reported in Table~\ref{tab:VFL}, down to 16 in 5 months. In 2 additional months, if the behaviour continues, the patient will decrease the VFL classification level from "very high" to "high" and in 7 more months will reach the status "normal". 

\begin{table}[htb]
    \centering
    \begin{tabular}{ccc}
 \toprule
Visceral Fat Level & Level Classification \\
 \midrule
1-9       & 0 (Normal)\\
10-14     & + (High)\\
15-30     & ++ (Very High)\\
\bottomrule
    \end{tabular}
    \\
    \caption{Visceral Fat Level reference table.}
    \label{tab:VFL}
\end{table}
%
%A similar trend analysis have been performed for all biological variables of the BIA exam. 

\begin{figure}[htb]
    \centering  
    \includegraphics[scale=0.55]{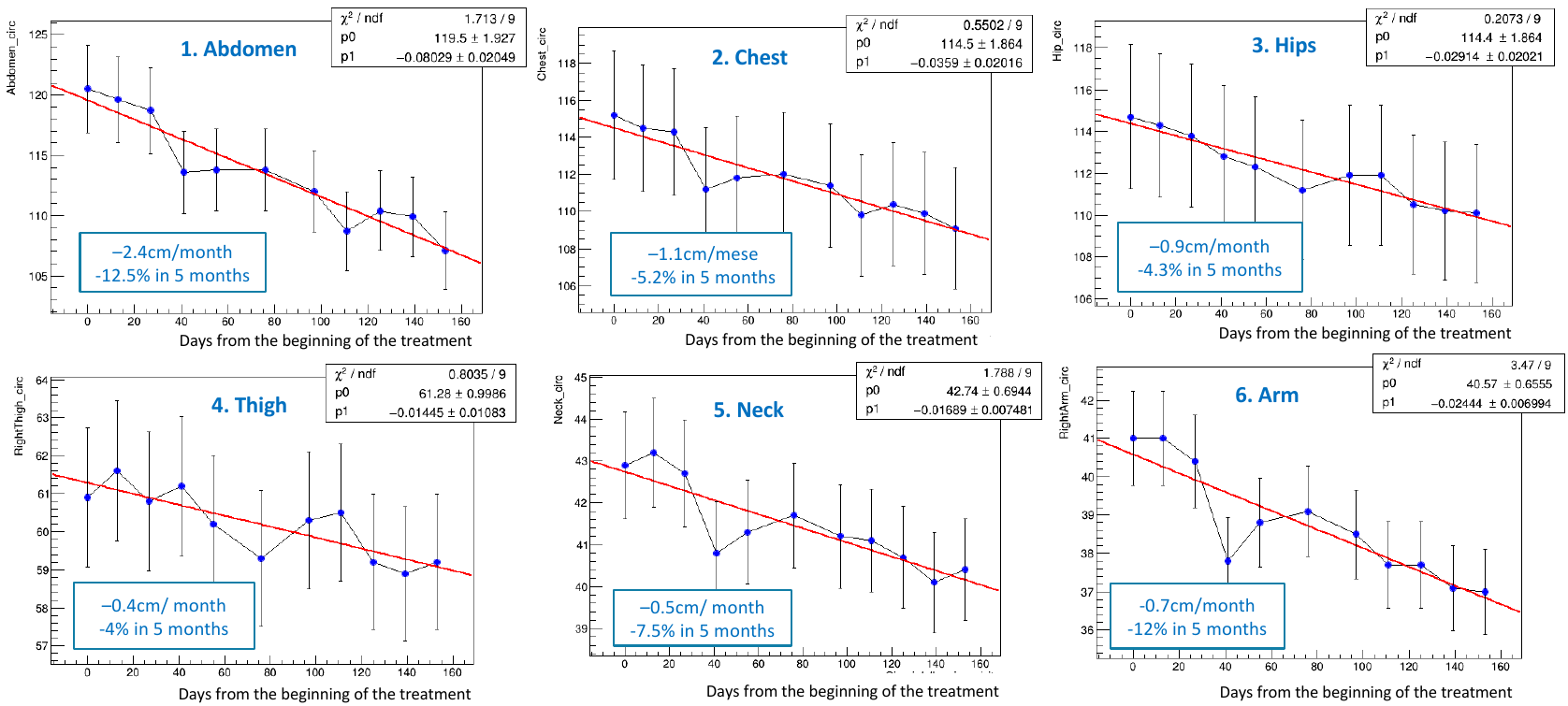}
    \caption{Trendlines of the body circumferences (Waist, Chest, Hips, Thigh, Neck, Arm) for the patient ID1556.  All numbers on the vertical axis are expressed in cm.
    %All circumferences drop linearly with time. Abdomen and Arm decreased of -12\% in 5 months, Chest, hips and thigh of about 4-5\%, neck of 7.5\%.
    }
    \label{fig:circum_ID1556}
\end{figure}

The study of the trend lines for the body circumferences is illustrated in Figure~\ref{fig:circum_ID1556}. The circumferences of waist, chest, hips, thigh, neck and arm are plotted as a function of the time from the beginning of the treatment. All circumferences drop linearly with time. Waist and arm circumferences decreased of 12\% in 5 months, chest, hips and thigh of about 4-5\%, neck of 7.5\%.

\begin{figure}[htb]
    \centering  
    \includegraphics[scale=0.55]{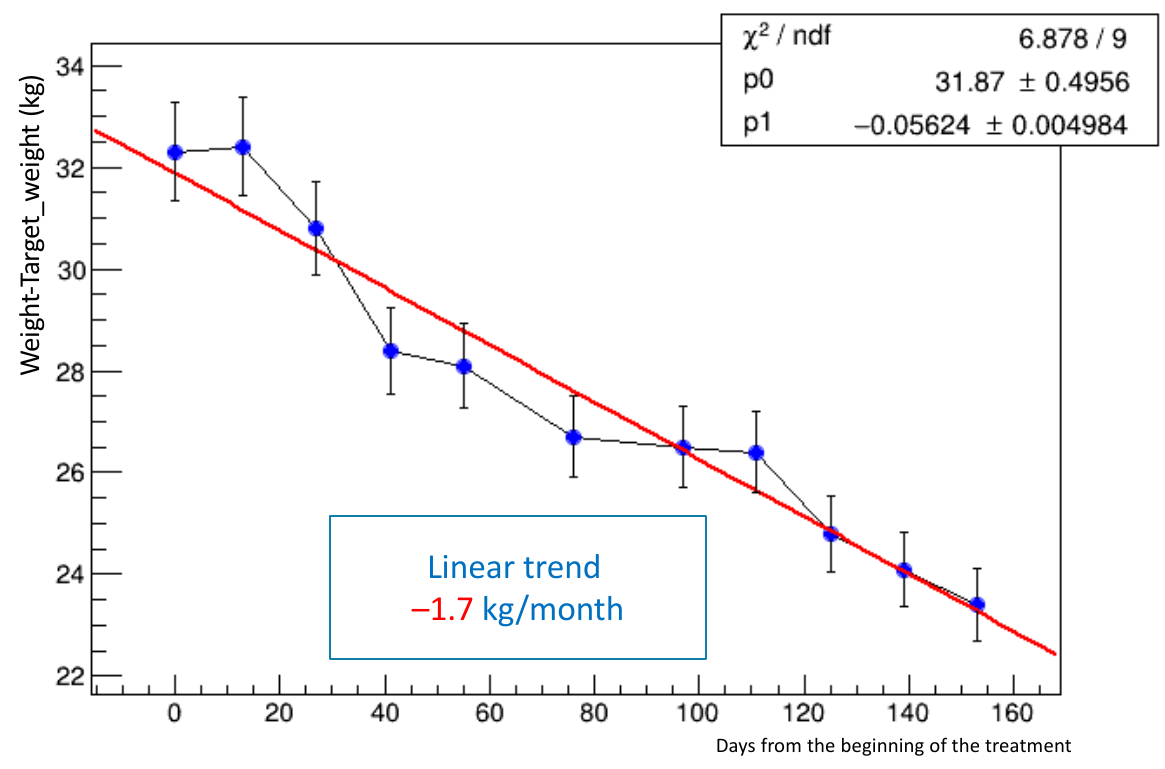}
    \caption{Trendline of the difference between the weight and the target weight (Weight-Target\_weight for the patient ID1556. 
    %This patient lost almost 2~kg (constantly) per month of treatment reaching a 30\% reduction of his weight in 5 months.}
    \label{fig:Target_ID1556}}
\end{figure}

Moreover, the study of the trend line for the difference between the weight and the target weight is shown in Figure~\ref{fig:Target_ID1556}. This patient lost almost 2~kg (constantly) per month of treatment reaching a 30\% reduction of his weight in 5 months.
%

%commento
%%%%%%%%%%%%%%%%%%%%%%%%%%%%%%%%%%%%%%
\subsection{Prediction of BIA parameters over time}
%%%%%%%%%%%%%%%%%%%%%%%%%%%%%%%%%%%%%%
\label{sec:pred_analysis} 

In order to generalize the time behaviour observed for several BIA variables for each patient from the repeated fit procedure, we inferred a general behaviour associated to each cluster by observing the distributions of the parameters of the trend line (a and b) calculated for every patient within each cluster. 
These values were used to fill two histograms per cluster, one for the initial value of the considered variable a and the other for the monthly variation of that parameter b*30 days. From these, the mean and the standard deviation of these distributions can be calculated, so that a prediction of the time behaviour can be done this time at cluster-level. \\

\begin{figure}[!htb]
    \centering  
    \includegraphics[width=.48 \textwidth]{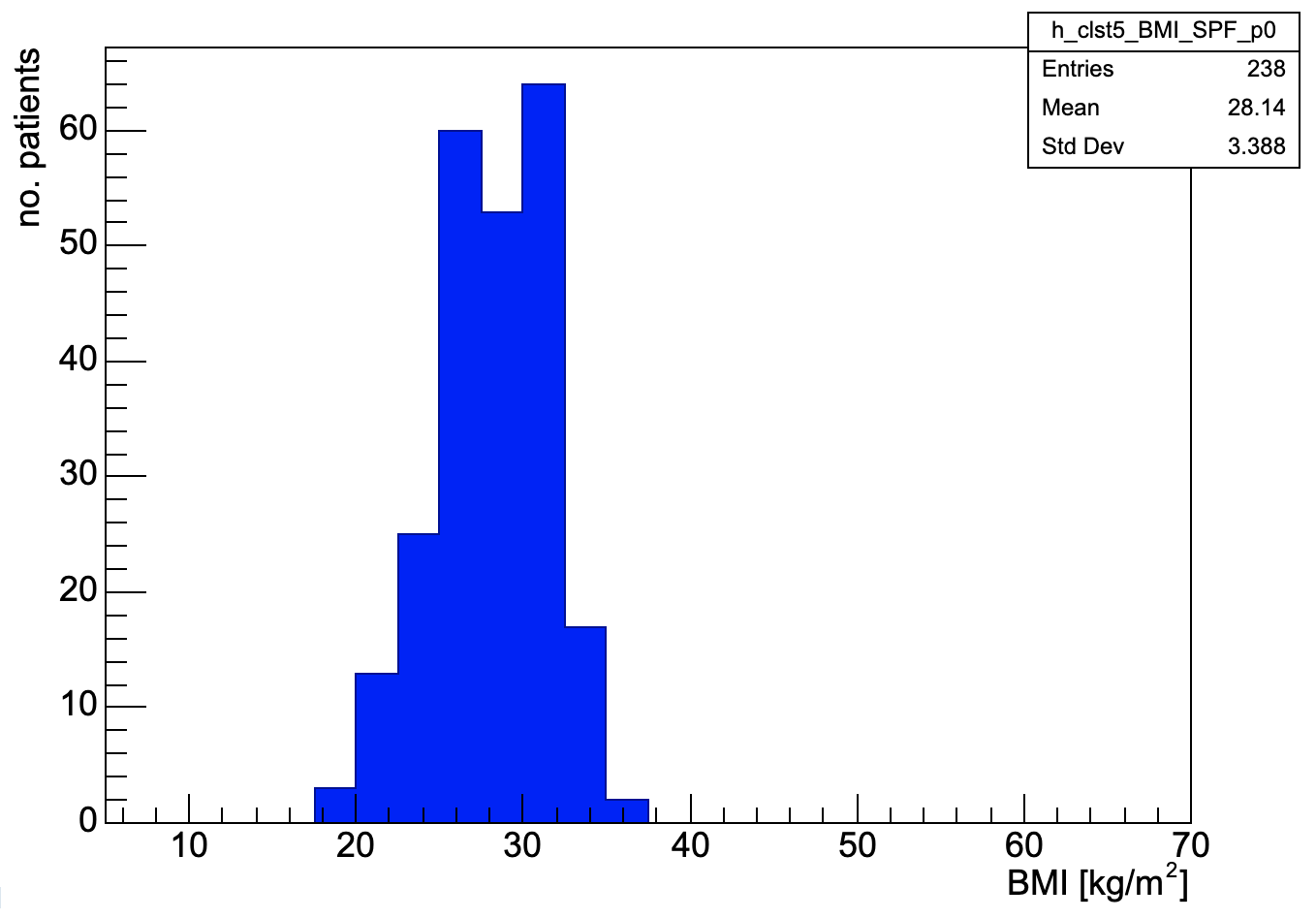}
    \includegraphics[width=.48 \textwidth]{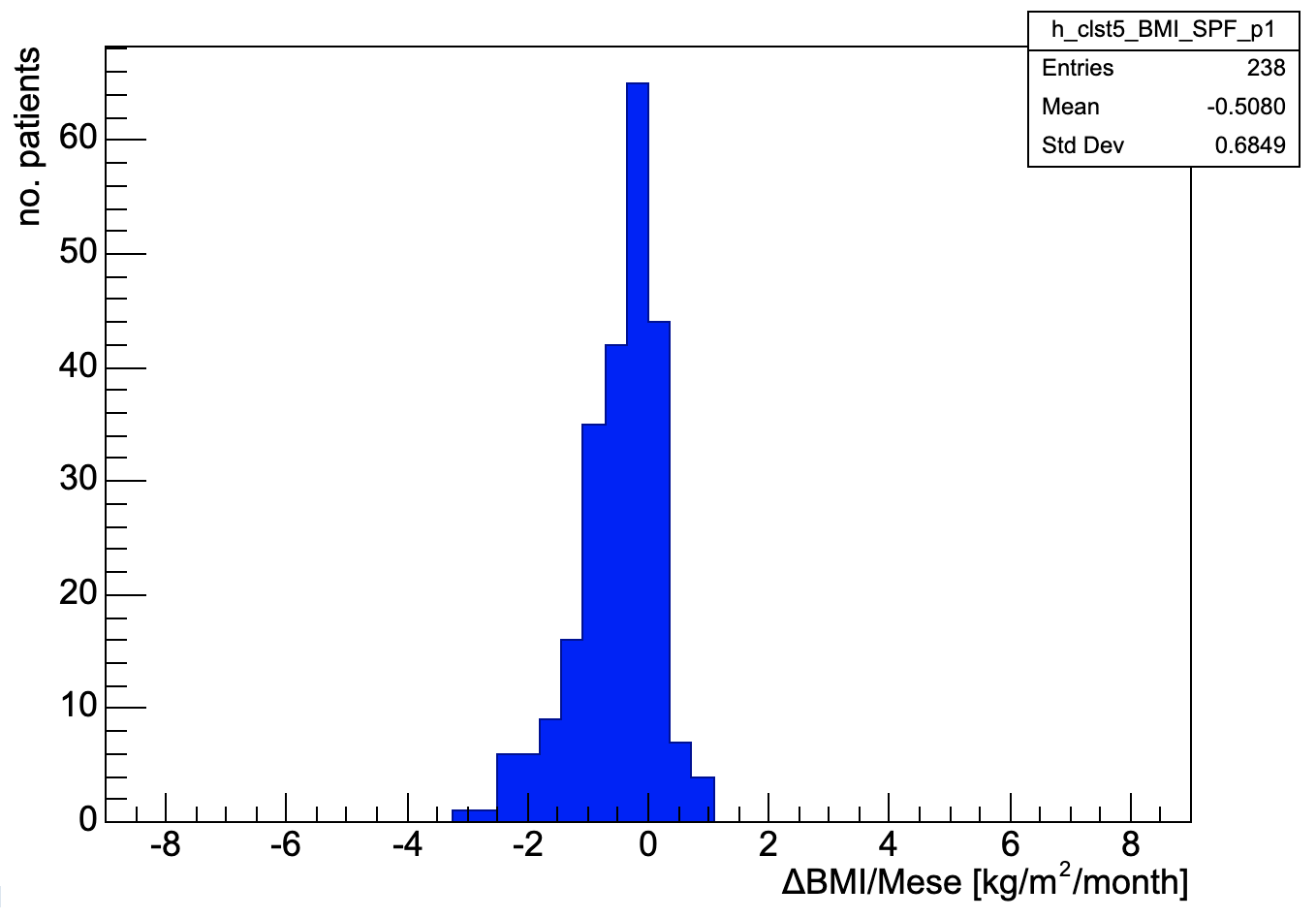}
    \caption{Distributions for patients in cluster 5 of the initial BMI values (on the left) and BMI variation per month (on the right) as obtained from the fit procedure.}
    \label{fig:BMI_dist}
\end{figure}

As an example, the results related to cluster 5 are reported. Cluster 5 is the one corresponding to male patients that have a BMI slightly above average. In Figure \ref{fig:BMI_dist} the distributions for the initial BMI and the BMI variation per month among all patients belonging to cluster 5 are shown. The second histogram contains the following information: if its mean value is negative, it means that, in average, these patients are decreasing their BMI value over the course of the diet, therefore losing weight; on the contrary, if its mean is positive, it means that they are increasing their BMI value, therefore gaining weight. In our case, the mean is about -0.5 BMI/month, which is an encouraging result since the cluster has an average initial value of 28.72 (Table \ref{tab:cluster_mean_val}). \\

An esteem of the number of months that would require for a patient in this cluster to reach a BMI value labelled as "Normal" can be obtained from the following equation:

\begin{equation}
\mathrm{N_{Months}} = \frac{(\mathrm{BMI_{ref}} - \mathrm{BMI_0})}{\Delta \mathrm{BMI/Month}}
\end{equation}

where $\mathrm{BMI_0}$ is the mean initial BMI for the cluster, $\mathrm{BMI_{ref}}$ is the target BMI and $\Delta \mathrm{BMI/Month}$ is the mean BMI variation per month. 
For this particular case, the target BMI value of 24.9 would be reached, in average, with about 7 months of diet.

Finally, to have a quantitative comparison between initial and final values of the BIA variables for each cluster, we considered their percentage variation over the course of the diet. Again, taking as an example the BMI, this is defined as

\begin{equation}
\mathrm{percentage~variation~BMI} = \frac{\mathrm{final~BMI} - \mathrm{initial~BMI}}{\mathrm{initial~BMI}}*100
\end{equation}

\begin{figure}[!htb]
    \centering  
    \includegraphics[width=.48 \textwidth]{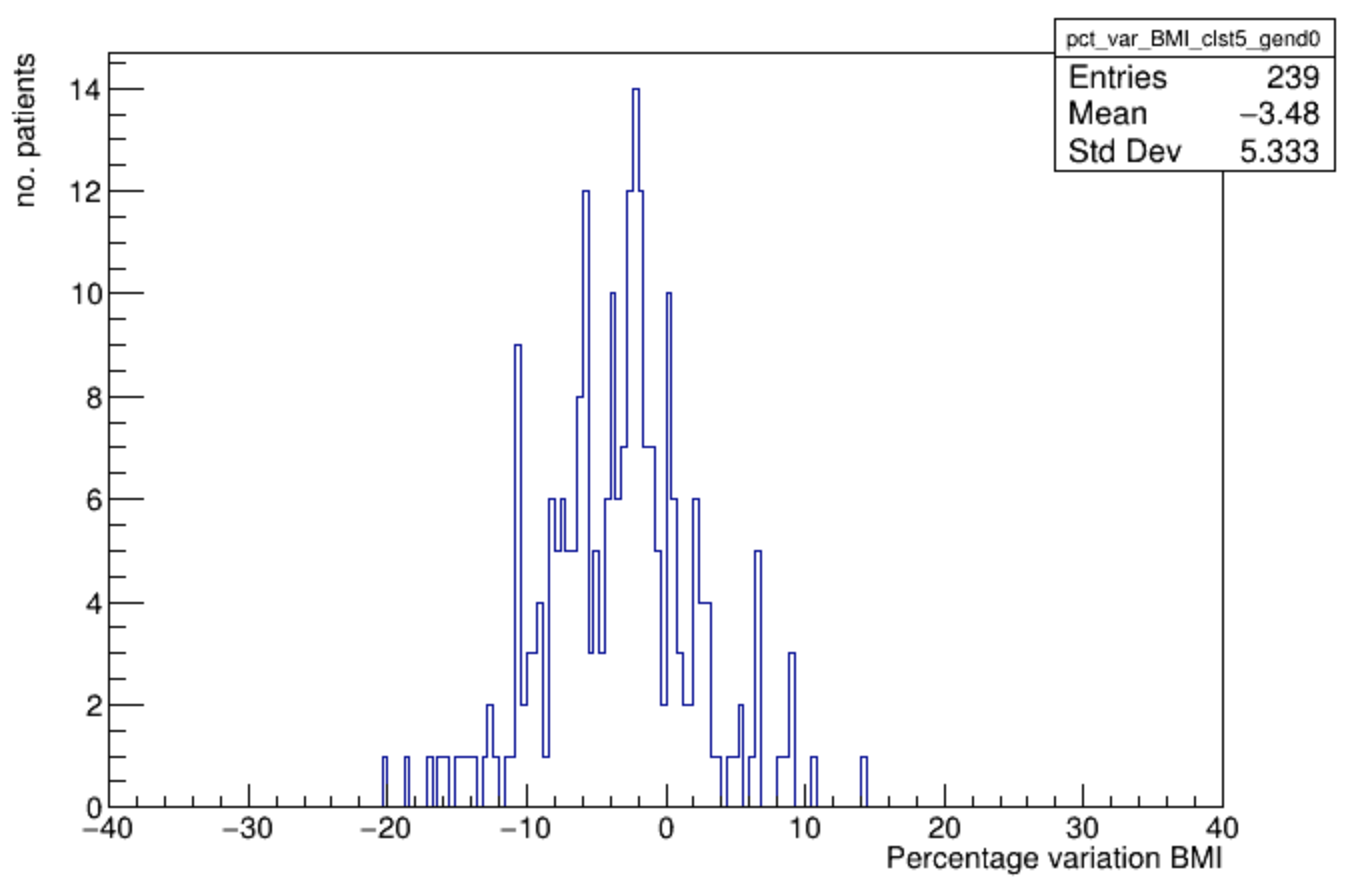}
    \includegraphics[width=.48 \textwidth]{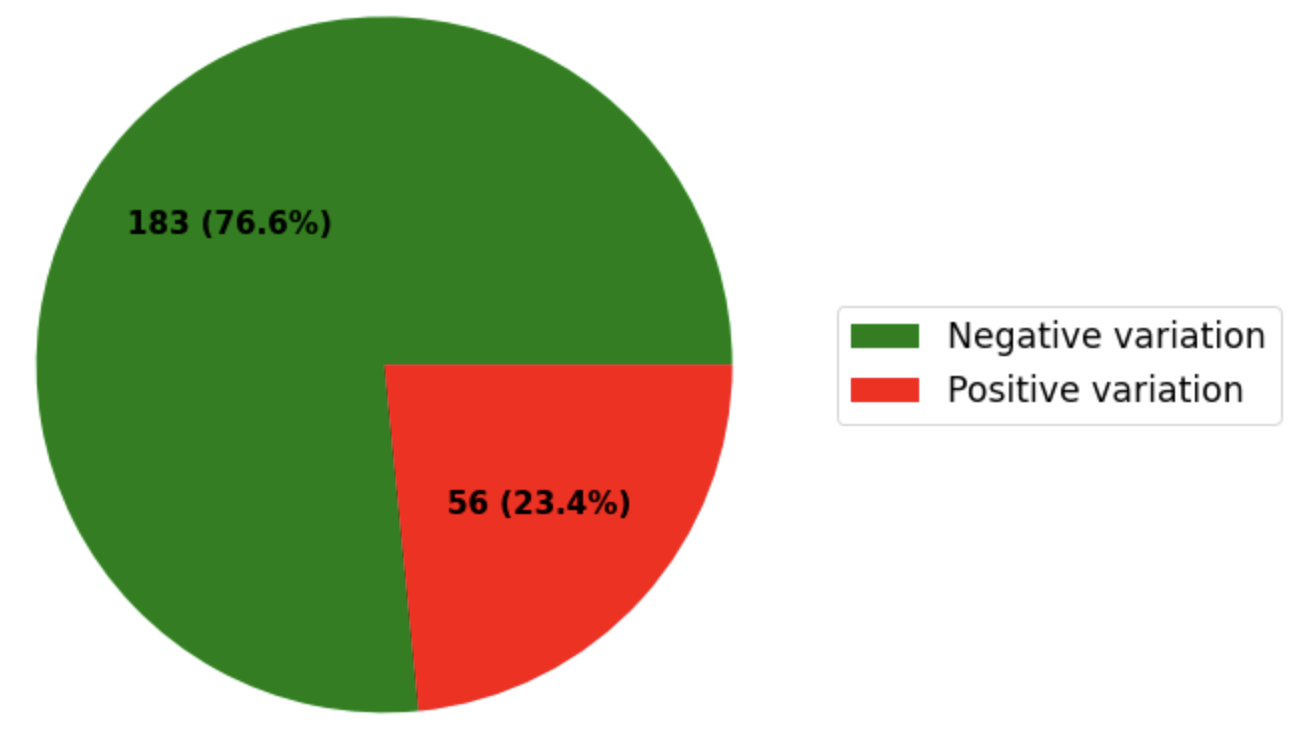}
    \caption{Distribution of the percentage variation of the BMI variable for cluster 5 patients and graph showing the fractions of negative and positive variations. More than 75\% of these subjects showed an improvement in their BMI toward the "Normal" reference values.}
    \label{fig:BMI_perc}
\end{figure}

Filling a histogram with these values for each patient in cluster 5 (Figure \ref{fig:BMI_perc}) allowed us to determine how many of them had a benefit from the diet, based on the number of negatives (if the objective is to decrease the considered variable) or positives (if the objective is to increase the considered variable) over the number of patients in the cluster. 

We can so see that about 77\% of subjects improved their health condition by decreasing their BMI thanks to the proposed dietary treatment. 

%%%%%%%%%%%%%%%%%%%%%%%%%%%%%%%%%%%%%%
\section{Results of the Data Analysis}
%%%%%%%%%%%%%%%%%%%%%%%%%%%%%%%%%%%%%%
The analysis described in Section~\ref{sec:time_analysis} and Section~\ref{sec:pred_analysis} is conducted for every patient and cluster. Trend lines are fitted to the selected variables for each patient within each cluster, and the mean values of these parameters are used to describe the trend for each cluster.

An overall quantitative comparison between initial and final values for each cluster is presented in Table~\ref{tab:BMI_final_comp}, detailing the changes in BMI, WHR, and TBW. Across clusters 1 to 9, a significant reduction in BMI is observed, while cluster 10, characterized as underweight, shows a discrete increase. Similarly, WHR decreases for all clusters except those classified as underweight, where it remains stable, and very severely obese clusters, where it increases.

A similar comparison is also produced for the circumference of chest, trunk and hip, as reported in Table~\ref{tab:circ_final_comp}. Also in this case a consistent reduction is observed for all the cluster from 1 to 9 and a moderate increase for the cluster 10.

\begin{table}[]
    \centering  
    \scalebox{0.55}{
\begin{tabular}{l |c c c c c  |c c c c c  |c c c c c }
 cluster  & \multicolumn{5}{|c}{BMI [kg/m$^2$]} & \multicolumn{5}{|c}{WHR} & \multicolumn{5}{|c}{TBW} \\
  & initial & final & $\Delta$[\%] & $\Delta^{i}_{target}$[\%] & $\Delta^{f}_{target}$[\%] & initial & final & $\Delta$[\%] & $\Delta^{i}_{target}$[\%] & $\Delta^{f}_{target}$[\%] & initial & final & $\Delta$[\%] & $\Delta^{i}_{target}$[\%] & $\Delta^{f}_{target}$[\%] \\
1 - M, Very severely obese & 55.80  & 48.25  & -13.5  & -55.4  & -48.4  & 1.09  & 1.12  & 3.5  & -17.1  & -19.8  & 36.83  & 40.83  & 10.9  & 68.3  & 51.8 \\
8 - M, Severely obese & 41.45  & 38.84  & -6.3  & -39.9  & -35.9  & 1.12  & 1.10  & -1.5  & -19.4  & -18.2  & 42.50  & 44.71  & 5.2  & 45.9  & 38.7 \\
5 - M, Overweight & 28.72  & 27.60  & -3.9  & -13.3  & -9.8  & 0.94  & 0.93  & -1.9  & -4.6  & -2.8  & 52.29  & 53.81  & 2.9  & 18.6  & 15.2 \\
9 - M, Moderately obese & 34.56  & 32.74  & -5.3  & -28.0  & -23.9  & 1.05  & 1.02  & -2.8  & -14.6  & -12.0  & 46.38  & 48.45  & 4.5  & 33.7  & 28.0 \\
7 - F, Normal weight, 35 yo & 26.04  & 25.29  & -2.9  & -4.4  & -1.5  & 0.90  & 0.89  & -1.3  & -11.3  & -10.2  & 48.06  & 49.41  & 2.8  & 20.7  & 13.3 \\
3 - F, Overweight, 55 yo & 30.66  & 29.20  & -4.8  & -18.8  & -14.7  & 0.97  & 0.95  & -2.1  & -17.8  & -16.0  & 41.64  & 43.60  & 4.7  & 39.3  & 28.5 \\
6 - F, Overweight, 27 yo & 33.62  & 32.32  & -3.9  & -25.9  & -23.0  & 0.99  & 0.97  & -1.3  & -18.8  & -17.7  & 40.03  & 41.46  & 3.6  & 44.9  & 35.1 \\
2 - F, Moderately obese, 44 yo & 38.83  & 36.63  & -5.7  & -35.9  & -32.0  & 1.04  & 1.02  & -1.9  & -22.9  & -21.3  & 37.51  & 39.15  & 4.4  & 54.6  & 43.0 \\
4 - Mix, Very severely obese & 48.07  & 44.84  & -6.7  & -48.2  & -44.5  & 1.04  & 1.07  & 2.1  & -18.6  & -20.2  & 35.42  & 36.63  & 3.4  & 69.4  & 63.8 \\
10 - Mix, Underweight & 18.82  & 19.09  & 1.5  & 0  & 0  & 0.79  & 0.79  & 0.1  & 0  & 0  & 58.24  & 58.21  & -0.0  & 0  & 0 \\
    \end{tabular}}
\\
    \caption{Initial and final values and percent difference for each cluster for the BMI, WHR, and TBW variables.}
    \label{tab:BMI_final_comp}
\end{table}

\begin{landscape}
\begin{table}[]
    \centering
    \resizebox{\textwidth}{!}{
    \begin{tabular}{l |c c c |c c c |c c c}
    Cluster & \multicolumn{3}{|c}{Chest circumference [cm]} & \multicolumn{3}{|c}{Trunk circumference [cm]} & \multicolumn{3}{|c}{Hip circumference [cm]} \\
    & initial & final & $\Delta$[\%]  & initial & final & $\Delta$[\%]  & initial & final & $\Delta$[\%] \\
    1 - M, Very severely obese & 130.30  & 128.47  & -1.4  & 153.53  & 147.38  & -4.0  & 141.35  & 132.10  & -6.5 \\
    8 - M, Severely obese & 123.75  & 120.73  & -2.4  & 137.94  & 132.41  & -4.0  & 123.68  & 120.39  & -2.7 \\
    5 - M, Overweight & 104.99  & 103.33  & -1.6  & 99.17  & 95.63  & -3.6  & 104.86  & 103.08  & -1.7 \\
    9 - M, Moderately obese & 115.22  & 112.82  & -2.1  & 120.62  & 114.68  & -4.9  & 114.55  & 111.91  & -2.3 \\
    7 - F, Normal weight, 35 yo & 93.92  & 92.79  & -1.2  & 88.40  & 86.20  & -2.5  & 97.86  & 96.68  & -1.2 \\
    3 - F, Overweight, 55 yo & 100.04  & 97.84  & -2.2  & 100.17  & 96.07  & -41  & 102.94  & 100.80  & -2.1 \\
    6 - F, Overweight, 27 yo & 105.87  & 103.97  & -1.8  & 107.73  & 104.47  & -3.0  & 109.35  & 107.43  & -1.8 \\
    2 - F, Moderately obese, 44 yo & 113.87  & 110.48  & -3.0  & 119.84  & 114.71  & -4.3  & 115.71  & 112.74  & -2.6 \\
    4 - Mix, Very severely obese & 126.27  & 122.36  & -3.1  & 134.27  & 132.13  & -1.6  & 128.84  & 124.26  & -3.6 \\
    10 - Mix, Underweight & 78.22  & 78.94  & 0.9  & 66.56  & 66.92  & 0.6  & 84.54  & 84.90  & 0.4 \\
    \end{tabular}}
    \\
    \caption{Initial and final values and percent difference for each cluster for the chest, trunk, and hip circumferences.}
    \label{tab:circ_final_comp}
\end{table}
\begin{table}[]
    \centering
\scalebox{0.6}{
\begin{tabular}{l |c c c c c c  |c c c c c c }
 cluster  & \multicolumn{6}{|c}{BMI [kg/m$^2$]} & \multicolumn{6}{|c}{WHR} \\
  & initial & $\Delta$/M & 6 months & 12 months & 18 months & months for target & initial & $\Delta$/M & 6 months & 12 months & 18 months & months for target \\
1 - M, Very severely obese & 54.16  & -1.4808  & 45.28  & 36.39  & 27.51  & 20  & 1.13  & 0.0024  & 1.14  & 1.16  & 1.17  & -\\
8 - M, Severely obese & 40.96  & -1.5404  & 31.72  & 22.48  & 13.24  & 11  & 1.11  & -0.0107  & 1.05  & 0.98  & 0.92  & 20 \\
5 - M, Overweight & 28.14  & -0.5080  & 25.09  & 22.05  & 19.00  & 7  & 0.93  & -0.0063  & 0.90  & 0.86  & 0.82  & 6 \\
9 - M, Moderately obese & 34.05  & -0.6825  & 29.96  & 25.86  & 21.77  & 14  & 1.05  & -0.0097  & 0.99  & 0.94  & 0.88  & 16 \\
7 - F, Normal weight, 35 yo & 25.77  & -0.3186  & 23.86  & 21.95  & 20.04  & 3  & 0.90  & -0.0041  & 0.88  & 0.85  & 0.83  & 25 \\
3 - F, Overweight, 55 yo & 30.35  & -0.4639  & 27.57  & 24.78  & 22.00  & 12  & 0.97  & -0.0003  & 0.97  & 0.97  & 0.96  & 518 \\
6 - F, Overweight, 27 yo & 33.19  & -0.5163  & 30.09  & 26.99  & 23.89  & 17  & 0.98  & -0.0026  & 0.97  & 0.95  & 0.94  & 72 \\
2 - F, Moderately obese, 44 yo & 38.46  & -0.7203  & 34.14  & 29.81  & 25.49  & 19  & 1.03  & -0.0071  & 0.99  & 0.95  & 0.91  & 33 \\
4 - Mix, Very severely obese & 47.23  & -0.8186  & 42.32  & 37.41  & 32.50  & 28  & 1.04  & 0.0039  & 1.07  & 1.09  & 1.11  & -\\
10 - Mix, Underweight & 18.85  & 0.0474  & 19.14  & 19.42  & 19.71  & 0  & 0.79  & 0.0105  & 0.85  & 0.91  & 0.97  & 0 \\
    \end{tabular}}
    \\
    \caption{Mean initial value, mean variation per month, predicted values at 6, 12 and 18 months and number of months to reach the target value for each cluster for BMI and WHR variables.}
    \label{tab:fixed_time}
\end{table}
\end{landscape}

%%%%%%%%%%%%%%%%%%%%%%%%%%%%%%%%%%%%%%
\section{Conclusion}
%%%%%%%%%%%%%%%%%%%%%%%%%%%%%%%%%%%%%%
This paper reports a statistical analysis of the performance of a low insulin index, alkaline and functional diet developed by ANTUR. The dataset was divided in clusters with similar characteristics evaluated with a Self-organizing map, a machine learning method. The sample of patients was checked on a regular basis with a BioImpedenziometric analysis using 40 features describing the global health status of the patient. A study of the behaviour of the BIA parameters evolution over time with respect to the first visit was performed. More than the 75\% of the patient that followed the ANTUR diet showed an improvement of the health status. The overall prediction for each cluster at fixed time periods, i.e. after 6, 12 and 18 months from the beginning of the diet, and the assumed number of months to reach the objective is reported in Table \ref{tab:fixed_time}, detailing the BMI and WHR parameters. Assuming a constant linear trend, most of these results show that values within the reference ranges can be obtained in a reasonable time, depending of course by the severity of the starting health condition of the patients.

The method proposed permits to have an estimation of time needed to a patient to arrive to the target weight and, more globally, to a healthier state. This could be a very useful tool alongside traditional ones to define a dedicated and more efficient protocol for each patient. Moreover, having a timeline can help to motivate the patient to continue in following the protocol.

%

%analyzed the biological parameters measured through BIA and develop a model to describe how they change over time since the start of the ANTUR treatment. By introducing a linear model and performing a least squares procedure, we can determine the initial values and daily variation rates of these parameters for each patient. Despite the varying goals of the different patient clusters, the functional low-insulin diet can be customized to meet their specific requirements. Through this analysis, we hope to uncover new insights that can improve our understanding of how the body responds to this treatment and ultimately lead to better outcomes for patients.

%\section*{Acknowledgments}
%This was supported in part by......

\section*{Acknowledgments}
This work was supported by Regione Campania that we thank for the funding. This work was also supported by the University of Naples Federico II, the University of Naples Pathenope, the Istituto Nazionale di Fisica Nucleare (INFN). We also thank the support staff from our institutions and the ANTUR Ricerca e Sviluppo s.r.l.%, which provided the data analyzed. We finally would like to thank all the patients that accepted to be included in the ANTUR diet program and in this research, without whom we could not have produced the results described in this paper. 

\clearpage
%Bibliography
\bibliographystyle{unsrt}  
\bibliography{ANTUR}  

\end{document}